\documentclass[onecolumn]{svjour3}         
\smartqed  
\usepackage{graphicx,amssymb,amsmath,epsf}
\usepackage{array,threeparttable,slashed}
\newcommand{\bp}{{\bf p}}

\newcommand{\bk}{{\bf k}}
\newcommand{\q}{{\bf q}}

\def\vz{{\mbox{\boldmath$0$}}}

\def\Spp{^3S_{1}^{+}}
\def\Dpp{^3D_{1}^{+}}

\begin{document}
\title{Covariant relativistic separable kernel approach for
electrodisintegration of the deuteron at high momentum transfer
}
\subtitle{Relativistic Description of Two- and Three-Body Systems
in Nuclear Physics, ECT*, October 13-19 2009}


\author{S.G. Bondarenko \and V.V. Burov \and
        E.P. Rogochaya
}


\institute{S.G. Bondarenko \and V.V. Burov \and E.P. Rogochaya \at
              Bogoliubov Laboratory of Theoretical Physics, Joint Institute for Nuclear Research, Dubna, Russia \\
           \and
              S.G. Bondarenko \\
              \email{bondarenko@jinr.ru}           
           \\
           ~~\\
           V.V. Burov \\
           \email{burov@theor.jinr.ru}
           \\
           ~~\\
           E.P. Rogochaya \\
              \email{rogoch@theor.jinr.ru}
}

\date{Received: date / Accepted: date}

\maketitle

\begin{abstract}
The paper considers the electrodisintegration of the deuteron for
kinematic conditions of the JLab experiment E-94-019. The
calculations have been performed within the covariant
Bethe-Salpeter approach with a separable kernel of nucleon-nucleon
interactions. The results have been obtained using the
relativistic plane wave impulse approximation and compared with
experimental data and other models. The influence of nucleon
electromagnetic form factors has been investigated.
\keywords{Separable ansatz \and Bethe-Salpeter equation \and
Deuteron electrodisintegration}
 \PACS{25.10.+s \and 25.30.Fj \and 11.10.St}
\end{abstract}
\section{Introduction}
\label{intro}
New experimental data for exclusive electrodisintegration of the
deuteron at high momentum transfer \cite{Egiyan:2007qj} can be a
good instrument for testing proposed relativistic models of
nucleon-nucleon (NN) interactions. The specific arrangement of the
experiment when the final state interaction (FSI) effects are
minimized allows one to compare results of the calculations
performed within the plane wave impulse approximation (PWIA).
Therefore, there is a chance to investigate the influence of
nucleon momentum distributions produced by various models
describing observables.

During last 15 years several relativistic models of NN
interactions have been elaborated
\cite{Laget:2004sm}-\cite{Arriaga:2007tc}. Generally, they are
based on fully relativistic expressions for matrix elements
depending on four-momenta of the nucleons under consideration.
However, in most cases there are difficulties caused by the
necessity to perform calculations with the zeroth component of
nucleon relative momentum $p_0$. They are solved by using some
constraints on $p_0$ obtained from physical assumptions that
limits the applicability of these models to low-energy
calculations. In practice, they are based on using nonrelativistic
nuclear interaction models (realistic
\cite{Machleidt:2000ge,Lacombe:1980dr} or separable potentials
\cite{Plessas:1982fs,Kwong:1997zz}). The attempt to apply the
covariant separable kernel \cite{Rupp:1989sg,Bondarenko:2002zz}
fails because of nonintegrable singularities which appear when
calculations in the high energy region are performed. This problem
was solved in \cite{Bondarenko:2008ha}-\cite{new} where the fully
relativistic covariant model was elaborated. Now it is interesting
to investigate how this model describes the electrodisintegration
of the deuteron at high momentum transfer where relativistic
effects are assumed to play an important role.

Another interesting problem is the influence of the used proton
and neutron electromagnetic form factors. The widely used model is
the dipole fit \cite{Pietschmann:1969mj}. However, it is well
known and intensively discussed that the relation of proton charge
form factor $G_{Ep}$ to the proton magnetic form factor $G_{Mp}$
obtained by the Rosenbluth separation technique, differs from the
one obtained by the recoil polarization method
\cite{Gayou:2001qd,Jones:1999rz}. To describe the results of the
latter method, it is necessary to use parametrization
\cite{Gayou:2001qd} for $G_{Ep}$. It should be noted that in this
case the Galster parametrization for the neutron electric form
factor $G_{En}$ \cite{Galster:1971kv} is applied
\cite{Laget:2004sm}. In this paper we compare the calculations
with the original dipole fit for the proton and neutron form
factors with those where the modified $G_{Ep}$ and $G_{En}$ are
used.

The paper is organized as follows. The formalism to describe NN
interactions within the Bethe-Salpeter approach with a separable
interaction kernel is presented in Sect.2. Sect.3 considers the
calculated cross section. The obtained results are discussed in
Sect.4.
\section{Formalism}
In the paper the deuteron electrodisintegraton is
considered within the Bethe-Salpeter (BS) approach
\cite{Salpeter:1951sz} with a separable kernel of NN
interactions. It is based on the solution of the BS equation:
\begin{eqnarray}
&&\Phi^{J{M}}(k;K)=
\frac{i}{(2\pi)^4} S_2(k; K) \int
{d^4p}V(k,p;K)\Phi^{J{M}}(p;K) \label{BS_Phi}
\end{eqnarray}
for the bound state of the neutron-proton ($np$) system with the
total angular momentum $J$ and its projection $M$ described by the
BS amplitude $\Phi^{J{M}}$. Here the total $K=k_p+k_n$ and the
relative $k=(k_p-k_n)/2$ momenta are used instead of the proton
$k_p$ and neutron $k_n$ momenta. In general, the BS amplitude can
be decomposed by the partial-wave states through the generalized
spherical harmonic ${\cal Y}$ and the radial part $\phi$
\cite{Bondarenko:2002zz} as:
\begin{eqnarray}
\Phi^{J{M}}_{\alpha\beta}(k; K_{(0)}) = \sum_{a} ({\cal
Y}_{aM}({\bk})U_C)_{\alpha\beta}\ \phi_{a}(k_0,|\bk|),
\label{phi00}
\end{eqnarray}
where $K_{(0)}=(M_d,\vz)$ is the total momentum of the NN system
in its rest frame (here it is the deuteron rest frame called the
laboratory system, LS); $M_d$ is the mass of the deuteron; $U_C$
is the charge conjugation matrix; $\alpha$, $\beta$ denote matrix
indices;  $a$ is a short notation of the partial-wave state
$^{2S+1}L_J^\rho$ with spin $S$, orbital $L$ and total $J$ angular
momenta, $\rho$ means positive- or negative-energy partial-wave
state. $S_2(k; K)$ is the free two-particle Green function:
$$S_2^{-1}(k; K)=\bigl(\tfrac12\:K\cdot\gamma+{k\cdot\gamma}-m\bigr)^{(1)}
\bigl(\tfrac12\:K\cdot\gamma-{k\cdot\gamma}-m\bigr)^{(2)}.$$ In
these calculations it is more convenient to use the BS vertex
function $\Gamma^{J{M}}$ which is connected with the BS amplitude
by the following relation:
\begin{eqnarray}
\Phi^{J{M}}(k;K)= S_2(k; K) \Gamma^{J{M}}(k;K).
\label{BS_vf}
\end{eqnarray}
After using the decomposition of type (\ref{phi00}) for the vertex
function the relation between the $\Phi^{J{M}}$ and $\Gamma^{J{M}}$
radial parts can be deduced:
\begin{eqnarray}
\phi_{a}(k_0,|\bk|)=\sum_{b}S_{ab}(k_0,|\bk|;s)g_{b}(k_0,|\bk|),
\label{amp_vf}
\end{eqnarray}
where $S_{ab}$ is the one-nucleon propagator
\cite{Bondarenko:2002zz}. To solve the BS equation (\ref{BS_Phi}),
we have used the separable ansatz for the interaction kernel
\begin{eqnarray}
V_{ab}(p_0, |\bp|; k_0, |\bk|; s)=
\sum_{i,j=1}^N\lambda_{ij}(s)
g_i^{[a]}(p_0, |\bp|)g_j^{[b]}(k_0, |\bk|), \label{V_separ}
\end{eqnarray}
where $N$ is a rank of the kernel, $g_i$ are model functions;
$\lambda$ is a parameter matrix satisfying the symmetry property
$\lambda_{ij}(s)=\lambda_{ji}(s)$; $k$ $[p]$ is the relative
momentum of the initial [final] nucleons; $s=(p_p+p_n)^2$ where
$p_p$ is the outgoing proton and $p_n$ is the neutron momentum,
respectively. If the radial part of the vertex function
$\Gamma^{JM}$ is written in the following form:
\begin{eqnarray}
g_a(p_0,|\bp|)=\sum_{i,j=1}^N{\lambda_{ij}(s)g_i^{[a]}(p_0,|\bp|)c_j(s)},
\label{phi_separ}
\end{eqnarray}
the initial integral BS equation (\ref{BS_Phi}) is transformed
into a system of linear homogeneous equations for the coefficients
$c_i(s)$:
\begin{eqnarray}
c_i(s)-\sum_{k,j=1}^N h_{ik}(s)\lambda_{kj}(s)c_{j}(s)=0,
\label{C_i}
\end{eqnarray}
where
\begin{eqnarray}
h_{ij}(s)=
-\frac{i}{4\pi^3}\sum_{a}\int dk_0\int \bk^2d|\bk|
\frac{g_i^{[a]}(k_0,|\bk|)g_j^{[a]}(k_0,|\bk|)}{(\sqrt
s/2-E_{\bk}+i\epsilon)^2-k_0^2} \label{H_separ}
\end{eqnarray}
and $E_\bk=\sqrt{\bk^2+m^2}$. Using (\ref{amp_vf}) and taking into
account only positive-energy partial-wave states for the deuteron
$^3S_1^+$, $^3D_1^+$, the radial part of the BS amplitude can be
written as follows:
\begin{eqnarray}
\phi_{a}(k_0,|\bk|)= \frac{g_{a}(k_0,|\bk|)}{({M_d}/2-E_{\bk}+i\epsilon)^2-k_0^2}.
\label{amp_vf_d}
\end{eqnarray}
Thus, using separable $g$ functions we can calculate observables
describing the $np$ system.
\section{Cross section}
When all particles are unpolarized the exclusive $d(e,e'n)p$
process can be described by the cross section in LS:
\begin{eqnarray}
&&\frac{d^3\sigma}{dQ^2d|\bp_n|d\Omega_n} =\frac{\sigma_{\rm Mott} \pi
\bp_n^2}{2(2\pi)^{3}M_dE_eE_e'}\nonumber \\
&&\times\left[ l^0_{00}W_{00}+l^0_{++}(W_{++}+W_{--})+ l^0_{+-}\cos2\phi~ 2{\rm Re}
W_{+-}\right.\nonumber\\
&&\hspace{4mm}- l^0_{+-}\sin2\phi~ 2{\rm Im}
W_{+-}-l^0_{0+} \cos\phi~ 2 {\rm Re} (W_{0+}-W_{0-})\nonumber\\
&&\hspace{3mm}\left.- l^0_{0+}\sin\phi~ 2{\rm Im} (W_{0+}+W_{0-})\right], \label{3cross_0}
\end{eqnarray}
where $\sigma_{\rm
Mott}=(\alpha\cos\frac{\theta}{2}/2E_e\sin^2\frac{\theta}{2})^2$
is the Mott cross section, $\alpha=e^2/4\pi$ is the fine structure
constant; $E_e$ [$E_e^\prime$] is the energy of the initial
[final] electron; $\Omega_e'$ is the outgoing electron solid
angle; $\theta$ is the electron scattering angle;
$Q^2=-q^2=-\omega^2+\q^2$, where $q=(\omega,\q)$ is the momentum
transfer. The outgoing neutron is described by momentum $\bp_n$
and solid angle $\Omega_n=(\theta_n,\phi)$ with zenithal angle
$\theta_n$ between $\q$ and $\bp_n$ momenta and azimuthal angle
$\phi$ between the $({\bf ee^\prime})$ and $(\q\bp_n)$ planes. The
photon density matrix elements have the following form:
\begin{eqnarray}
&&l_{00}^0=\frac{Q^2}{\q^2},\quad l_{0+}^0=\frac{Q}{|\q|\sqrt
2}\sqrt{\frac{Q^2}{\q^2}+\tan^2 \frac{\theta}{2}},\quad\nonumber\\
&&l_{++}^0=\tan^2\frac{\theta}{2}+\frac{Q^2}{2\q^2},\quad
l_{+-}^0=-\frac{Q^2}{2\q^2}.
\end{eqnarray}
The hadron density matrix elements
\begin{eqnarray}
W_{\lambda\lambda'}=W_{\mu\nu}\varepsilon_\lambda^\mu{\varepsilon_{\lambda'}}^\nu,
\end{eqnarray}
where $\lambda$, $\lambda'$ are photon helicity components
\cite{Dmitrasinovic:1989bf}, can be calculated using the photon
polarization vectors $\varepsilon$ and Cartesian components of
hadron tensor
\begin{eqnarray}
W_{\mu\nu}=\frac{1}{3}\sum_{s_ds_ns_p}\left|<np:SM_S|j_\mu|d:1M>\right|^2,
\label{ht}
\end{eqnarray}
where $S$ is the spin of the $np$ pair and $M_S$ is its
projection. The hadron current $j_\mu$ in (\ref{ht}) can be written
according to the Mandelstam technique \cite{Mandelstam:1955sd} and
has the following form:
\begin{eqnarray}
&&<np:SM_S|j_\mu|d:1M>=i\sum_{r=1,2}\int\frac{d^4p}{(2\pi)^4}{\rm
Sp}\left\{\Lambda({\cal L
}^{-1})\bar\psi_{SM_S}(p^{\rm {CM}};P^{\rm {CM}})
\Lambda({\cal L
})\right.\nonumber\\
&&\left. \times\Gamma_\mu^{(r)}(q)S^{(r)}\left(\frac{K_{(0)}}{2}-(-1)^rp-\frac
q2\right)\Gamma^M
\left(p+(-1)^r\frac{q}{2};K_{(0)}\right)\right\} \label{cur_fsi}
\end{eqnarray}
within the relativistic impulse approximation. The sum over
$r=1,2$ corresponds to the interaction of the virtual photon with
the proton and with the neutron in the deuteron, respectively.
Total $P^{\rm CM}$ and relative $p^{\rm {CM}}$ momenta of the
outgoing nucleons are considered in the final $np$ pair rest frame
(center-of-mass system, CM) and can be written in LS using the
Lorenz-boost transformation along the $\q$ direction. The Lorenz
transformation of $np$ pair wave function $\psi_{SM_S}$ from CM to
LS is:
\begin{eqnarray}
\Lambda
({\cal L})=&&\left(\frac{1+\sqrt{1+\eta}}{2}\right)^{\frac{1}{2}}
\left(1+\frac{\sqrt{\eta}\gamma_0\gamma_3}{1+\sqrt{1+\eta}}\right),
\end{eqnarray}
where $\eta=\q^2/s$. 
The
interaction vertex is chosen in the on-mass-shell form:
\begin{eqnarray}
\Gamma_\mu(q)=\gamma_\mu F_1(q^2)-\frac{1}{4m}\left(\gamma_\mu\slashed{q}-\slashed{q}\gamma_\mu\right)F_2(q^2),\label{intervertex}
\end{eqnarray}
here $F_1(q^2)$ is the Dirac form factor, $F_2(q^2)$ - Pauli form
factor. The form factors are described by the dipole fit model
\cite{Pietschmann:1969mj} or modified dipole fit
\cite{Gayou:2001qd,Galster:1971kv}. If the outgoing nucleons are
supposed to be non-interacting then this is the so-called plane-wave
approximation. In this case the $np$ pair wave function can be
written in the following form:
\begin{eqnarray}
&&\bar\psi_{SM_S}(p;P)~\rightarrow
\bar\psi_{SM_S}^{(0)}(p,{p^*};P)= (2\pi)^4\bar\chi_{SM_S}(p;P)\delta(p-p^*),\label{wf_PWA}
\end{eqnarray}
where $p^*=(0,\bp^*)$ is the relative momentum of on-mass-shell
nucleons, $\chi_{SM_S}$ describes spinor states of the pair. Taking into
account representation (\ref{wf_PWA}), the hadron current
(\ref{cur_fsi}) can be transformed into a sum:
\begin{eqnarray}
&&<np:SM_S|j_\mu|d:1M>=i\sum_{r=1,2}\left\{\Lambda({\cal
L}^{-1})\bar\chi_{SM_S}\left({p^*}^{\rm CM}; P^{\rm CM}\right)\Lambda({\cal
L})\Gamma_\mu^{(r)}(q) \right.\nonumber\\
&&\hspace{23mm}\times S^{(r)}\left(\frac{K_{(0)}}{2}-(-1)^r
p^*-\frac{q}{2}\right)\left.\Gamma^{M}\left(p^*+(-1)^r\frac{q}{2}; K_{(0)}\right)\right\}.\label{currentcalc}
\end{eqnarray}

In this paper the cross section of the exclusive
electrodisintegration of the deu\-te\-ron
${d^2\sigma}/{dQ^2d|\bp_n|}$ \cite{Egiyan:2007qj} is calculated.
It can be obtained from (\ref{3cross_0}) after integration over
the neutron solid angle:
\begin{eqnarray}
\frac{d^2\sigma}{dQ^2d|\bp_n|}=\int\limits_{\Omega_n}
\frac{d^3\sigma}{dQ^2d|\bp_n|d\Omega_n}d\Omega_n. \label{cs_jlab}
\end{eqnarray}
According to \cite{Egiyan:2007qj} the integration is performed over
$\Omega_n$: $20^\circ\leqslant\theta_n\leqslant 160^\circ$,
$0^\circ\leqslant\phi\leqslant 360^\circ$. Four different $Q^2$
are considered. The obtained results are discussed in the next
section.
\section{Results and discussion}
In this paper the exclusive cross section of electrodisintegration
(\ref{cs_jlab}) for kinematic conditions of the JLab experiment
\cite{Egiyan:2007qj} has been calculated within the Bethe-Salpeter
approach with the rank-six separable kernel MY6 \cite{new}. The
calculations have been performed within the relativistic PWIA. The
obtained results have been compared with experimental data and two
theoretical models, the nonrelativistic Graz II (NR)
\cite{Mathelitsch:1981mr} and relativistic Graz II
\cite{Rupp:1989sg} separable interaction kernels.

\begin{figure}
\begin{center}
  \includegraphics[width=0.55\textwidth]{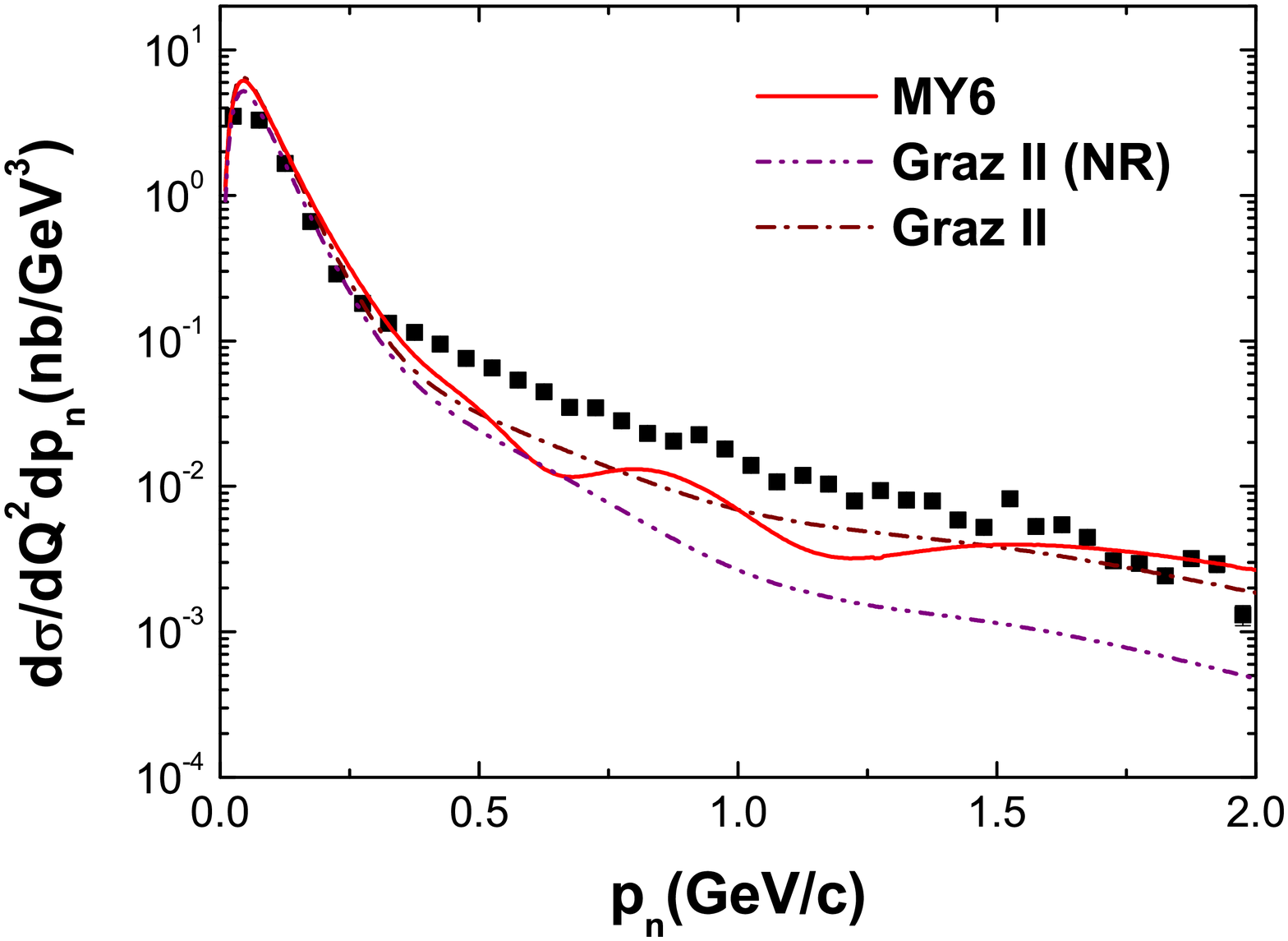}
\end{center}  
\caption{The cross section (\ref{cs_jlab}) depending on neutron momentum $\bp_n$ is considered for
$Q^2=2\pm0.25$\,GeV$^2$. Calculations with the Graz II (NR)
\cite{Mathelitsch:1981mr} (purple dash-dot-dotted line), Graz II
\cite{Rupp:1989sg} (brown dash-dotted line) and MY6 \cite{new}
(red solid line) models are present. The dipole fit model
\cite{Pietschmann:1969mj} for nucleon form factors is used. }
\label{fig:1}
\end{figure}

\begin{figure}
\begin{center}
  \includegraphics[width=0.55\textwidth]{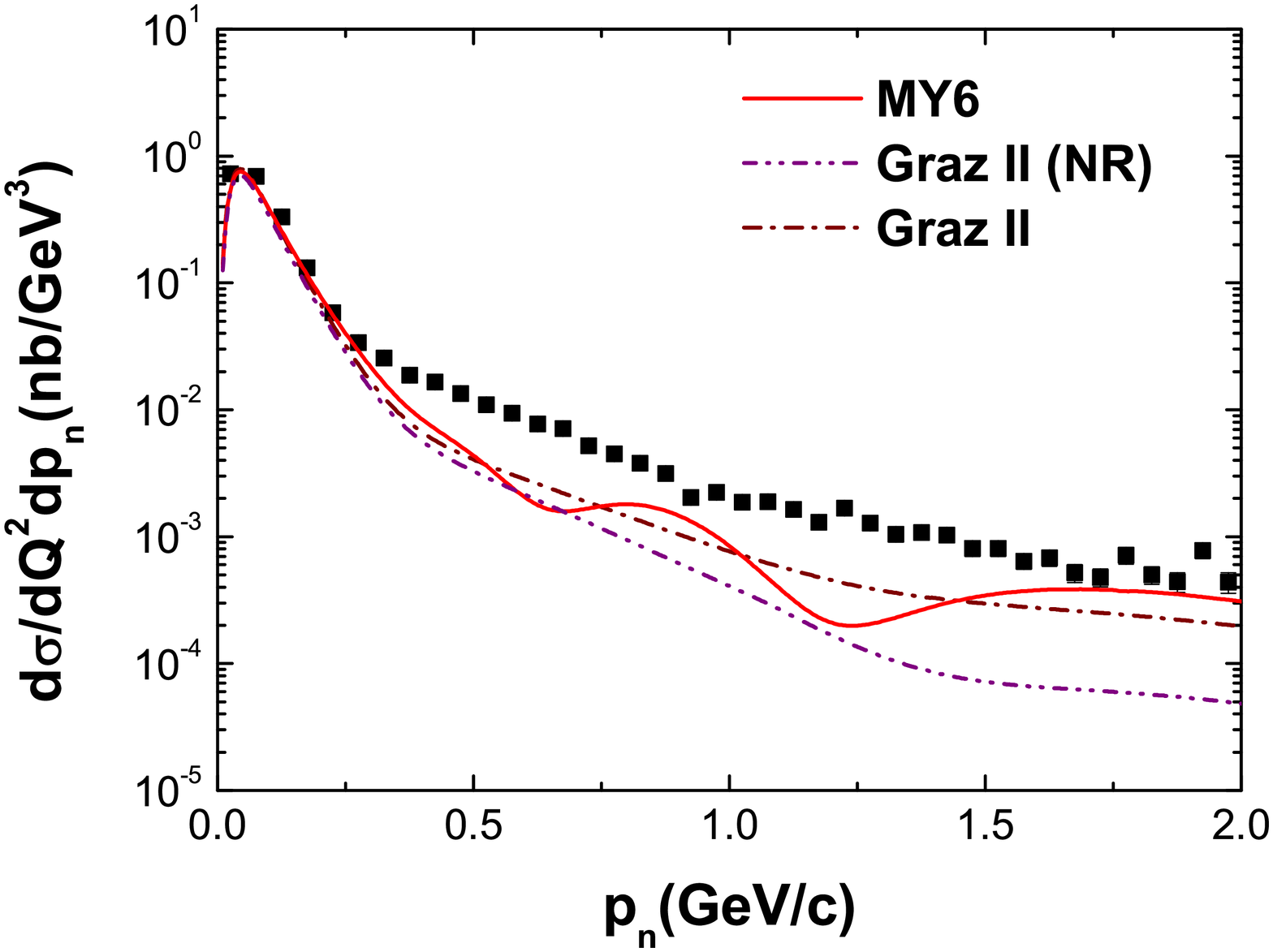}
\end{center}
\caption{As in Fig.\ref{fig:1}, but for $Q^2=3\pm0.5$\,GeV$^2$.} \label{fig:2}
\end{figure}

\begin{figure}
\begin{center}
  \includegraphics[width=0.55\textwidth]{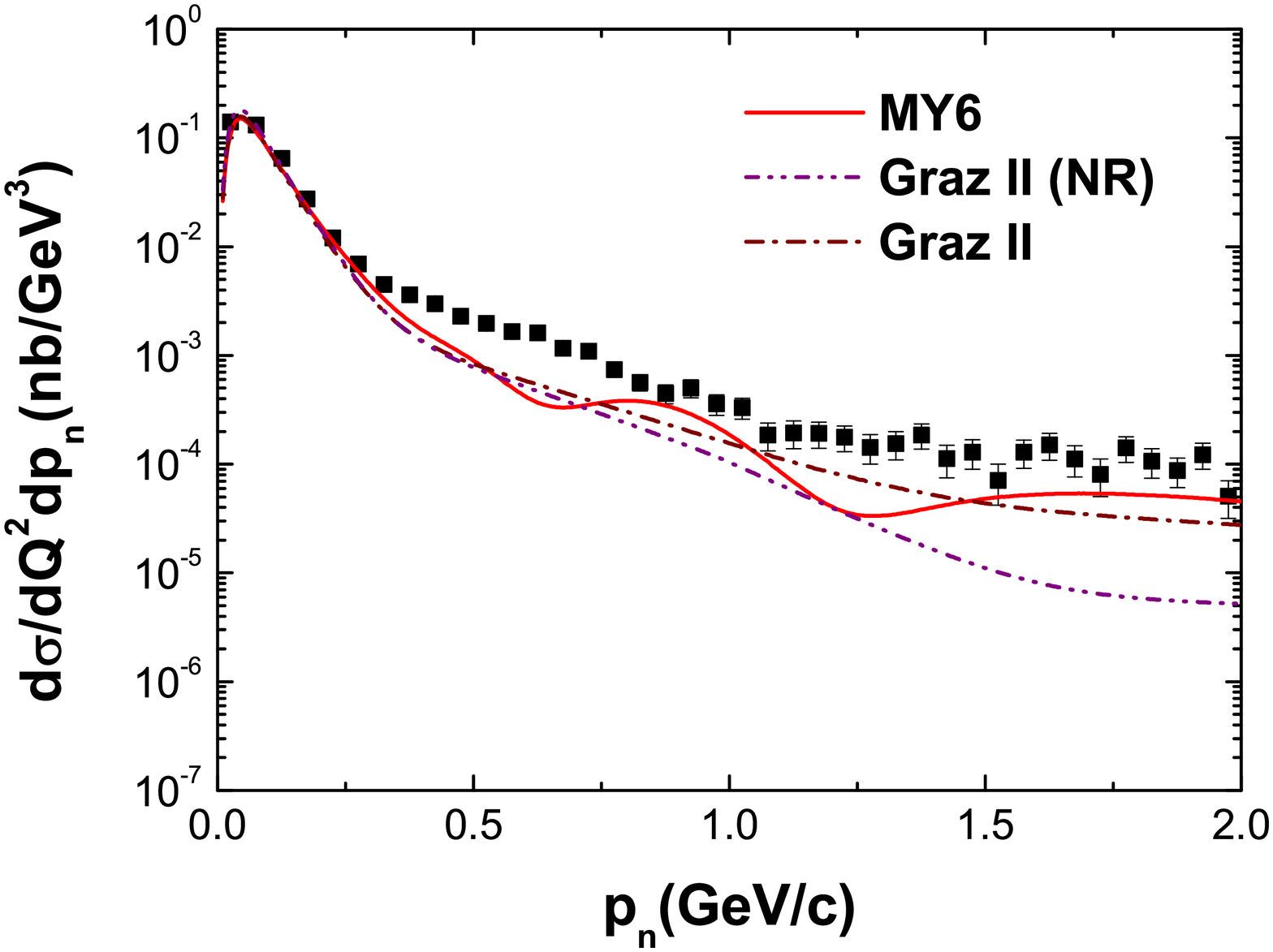}
\end{center}
\caption{As in Fig.\ref{fig:1}, but for $Q^2=4\pm0.5$\,GeV$^2$.} \label{fig:3}
\end{figure}

\begin{figure}
\begin{center}
  \includegraphics[width=0.55\textwidth]{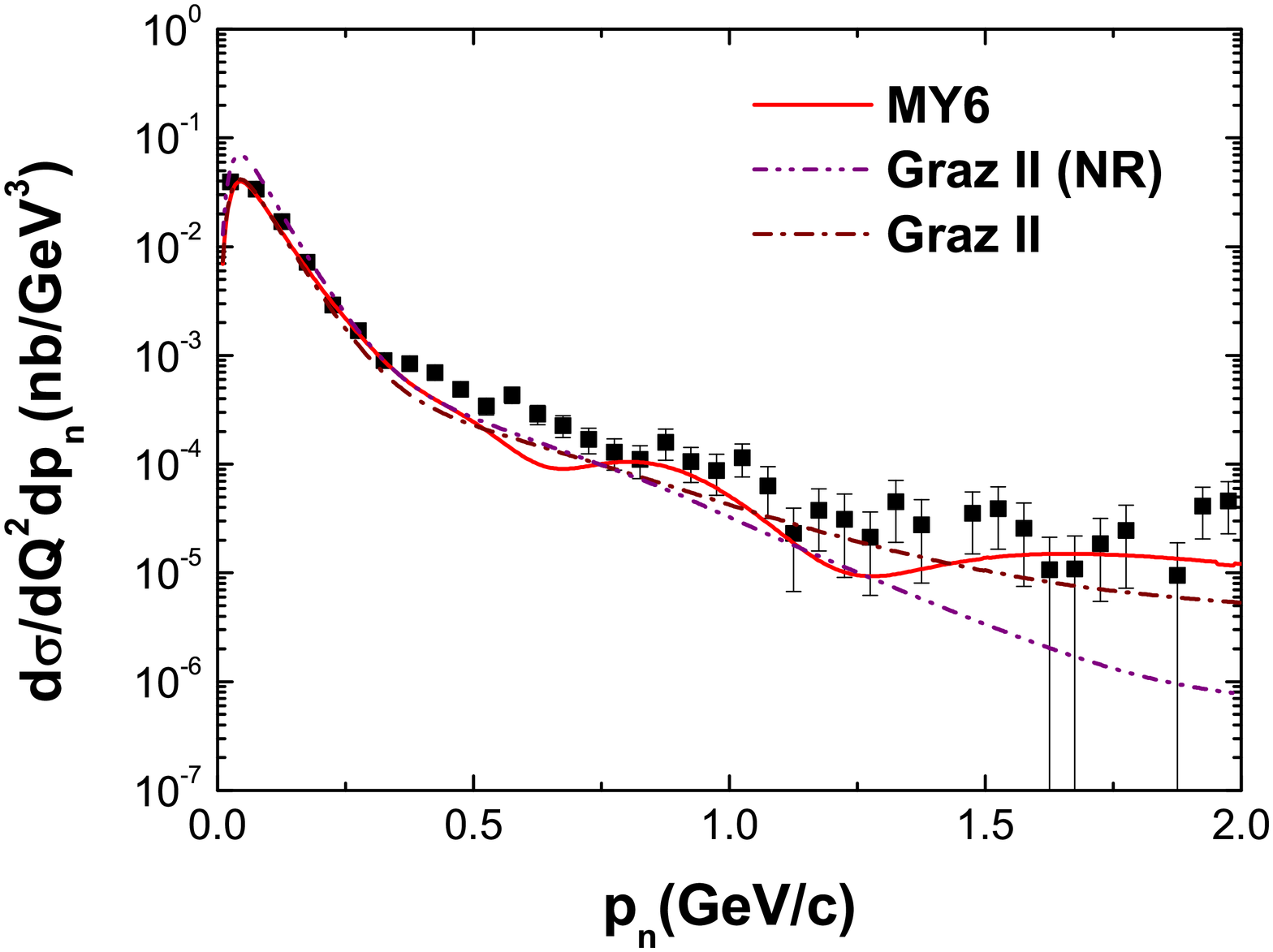}
\end{center}
\caption{As in Fig.\ref{fig:1}, but for $Q^2=5\pm0.5$\,GeV$^2$.} \label{fig:4}
\end{figure}

Figs. \ref{fig:1}-\ref{fig:4} illustrate the cross section
depending on outgoing neutron momentum $\bp_n$ for $Q^2$=2, 3, 4,
5\,GeV$^2$, respectively. The dipole fit model
\cite{Pietschmann:1969mj} for the nucleon electromagnetic form
factors has been used. One nonrelativistic Graz II (NR) and two
relativistic MY6, Graz II separable kernels of NN interactions
have been investigated. A good agreement with the experimental
data can be seen at low neutron momenta $|\bp_n|<$0.25\,GeV/c on
the figures. The discrepancy between the theoretical models and
the experimental data increases with $|\bp_n|>$0.25\,GeV/c for all
the considered models. However, we see the agreement of the
relativistic models (MY6, Graz II) with the experimental data at
high neutron momenta. Moreover, the relativistic description
becomes better with $Q^2$ increasing and theoretical curves go
practically along experimental points at $Q^2=5$\,GeV$^2$.
Therefore, relativistic effects play an important role in
description of the deuteron electrodisintegration at high momentum
transfer and high neutron momenta.

\begin{figure}
\begin{center}
  \includegraphics[width=0.55\textwidth]{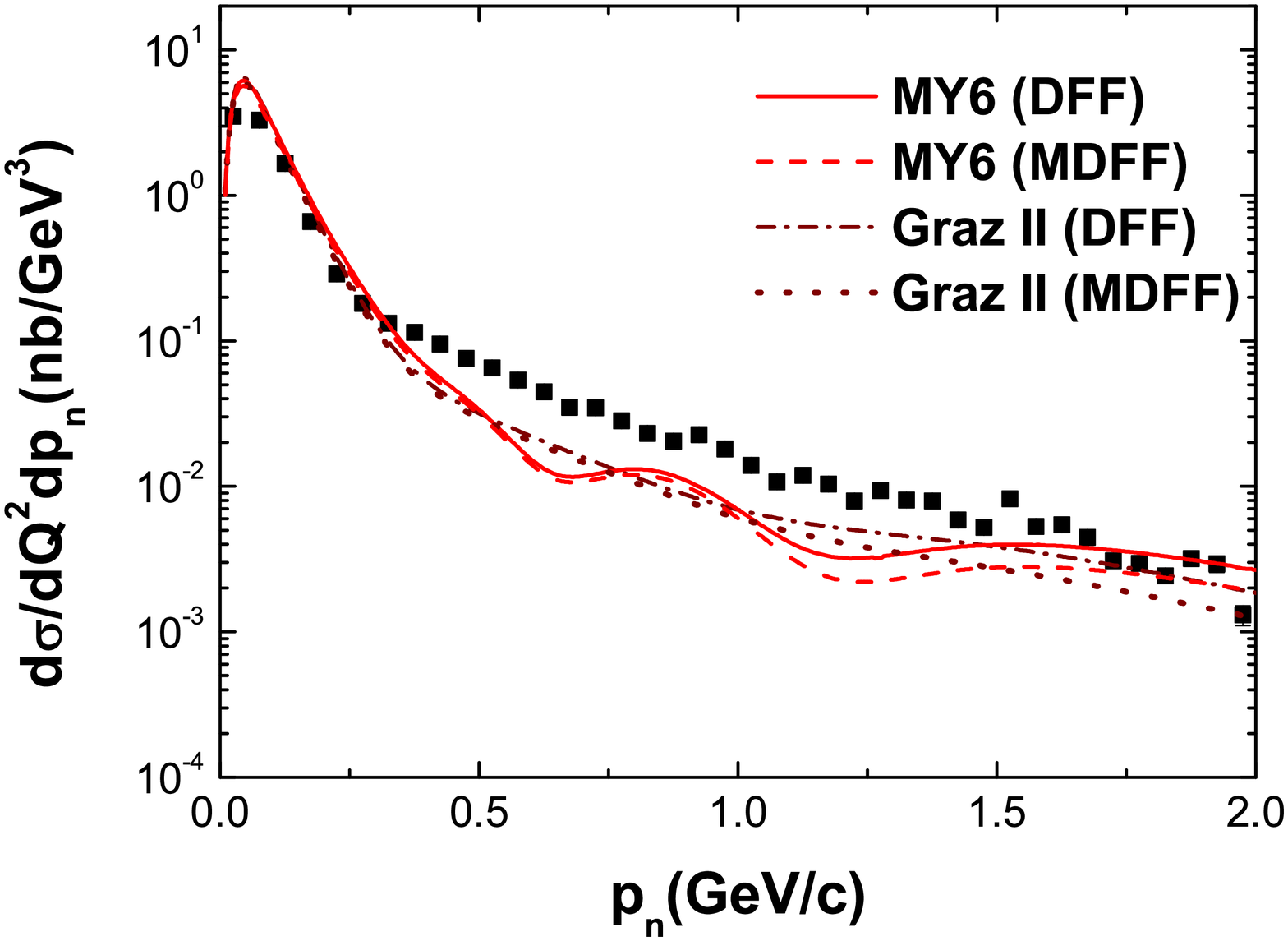}
\end{center}
\caption{The calculated cross sections (\ref{cs_jlab}) by using the
dipole fit model for nucleon electromagnetic form factors
\cite{Pietschmann:1969mj} (MY6 (DFF) - red solid line, Graz II
(DFF) - brown dash-dotted line) are compared to those obtained
with modified $G_{Ep}$ \cite{Gayou:2001qd} and $G_{En}$
\cite{Galster:1971kv} (MY6 (MDFF) - red dashed line, Graz II
(MDFF) - brown dotted line). $Q^2=2\pm0.5$\,GeV$^2$.}
\label{fig:5}
\end{figure}

\begin{figure}
\begin{center}
  \includegraphics[width=0.55\textwidth]{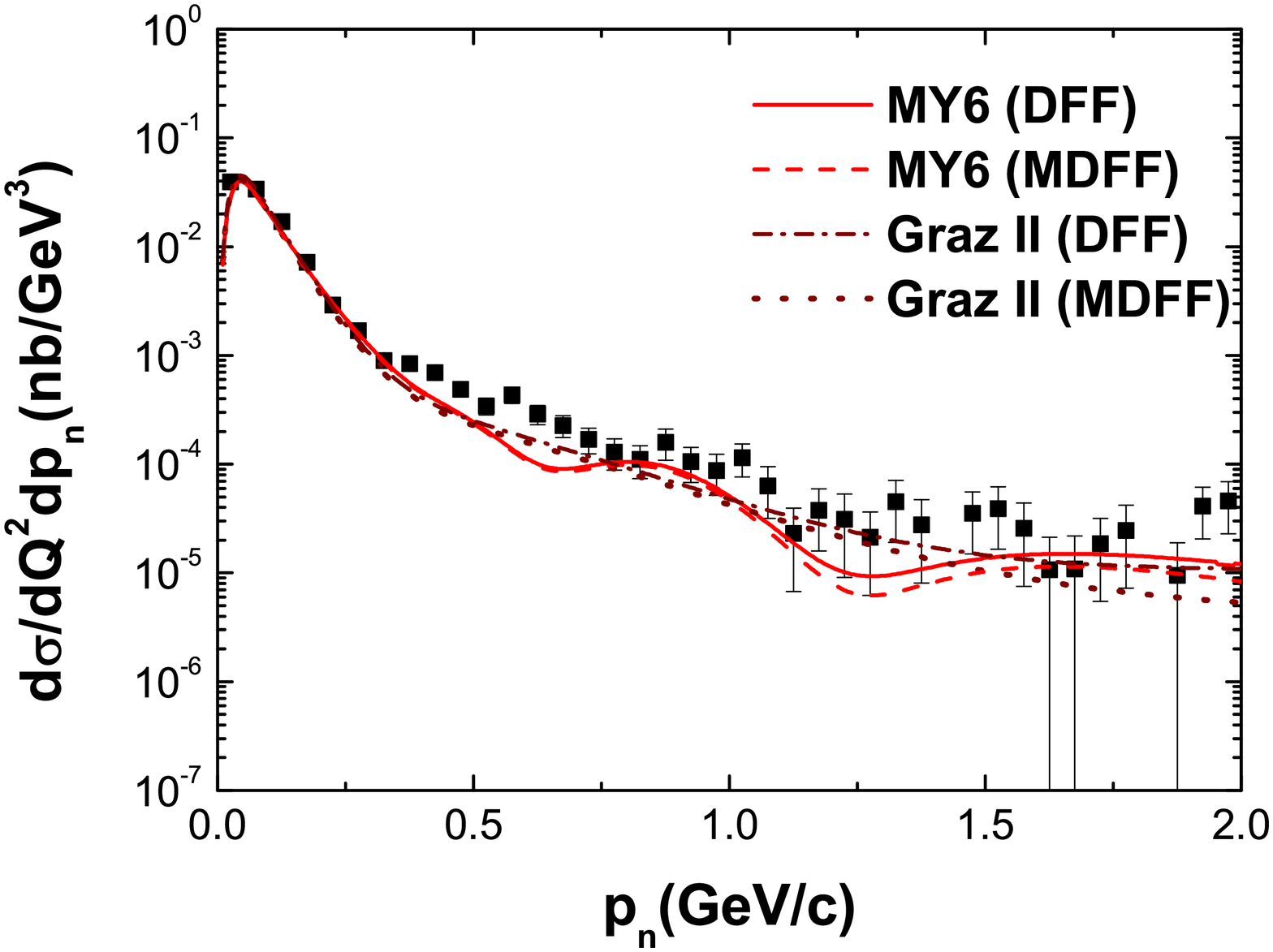}
\end{center}
\caption{As in Fig.\ref{fig:5}, but for $Q^2=5\pm0.5$\,GeV$^2$.}
\label{fig:6}
\end{figure}

The calculations with the modified proton $G_{Ep}$
\cite{Gayou:2001qd} and neutron $G_{En}$ \cite{Galster:1971kv}
form factors at $Q^2$=2\,GeV$^2$ (Fig. \ref{fig:5}) and
$Q^2$=5\,GeV$^2$ (Fig. \ref{fig:6}) are compared to those obtained
using the dipole fit model for nucleon form factors. Two
relativistic models of NN interactions MY6 and Graz II have been
considered. All the theoretical calculations agree with the experiment
at $|\bp_n|<0.25$\,GeV/c and begin to deviate from it with $\bp_n$
increasing. We can also see slight difference between the
cross sections obtained using the dipole fit and modified dipole
fit models for the nucleon electromagnetic form factors.
It is interesting that
the results calculated within the dipole fit model, which does not
describe the behavior of the electric form factor of the proton at
high $Q^2$, are virtually undistinguishable from those obtained
with the modified $G_{Ep}$ \cite{Gayou:2001qd}. However, the final
conclusion which model is better can be made only when the final state
interactions,
negative-energy partial-wave states ($P$ waves) and two-body currents (TBC) are taken into account.

\begin{figure}
\begin{center}
  \includegraphics[width=0.55\textwidth]{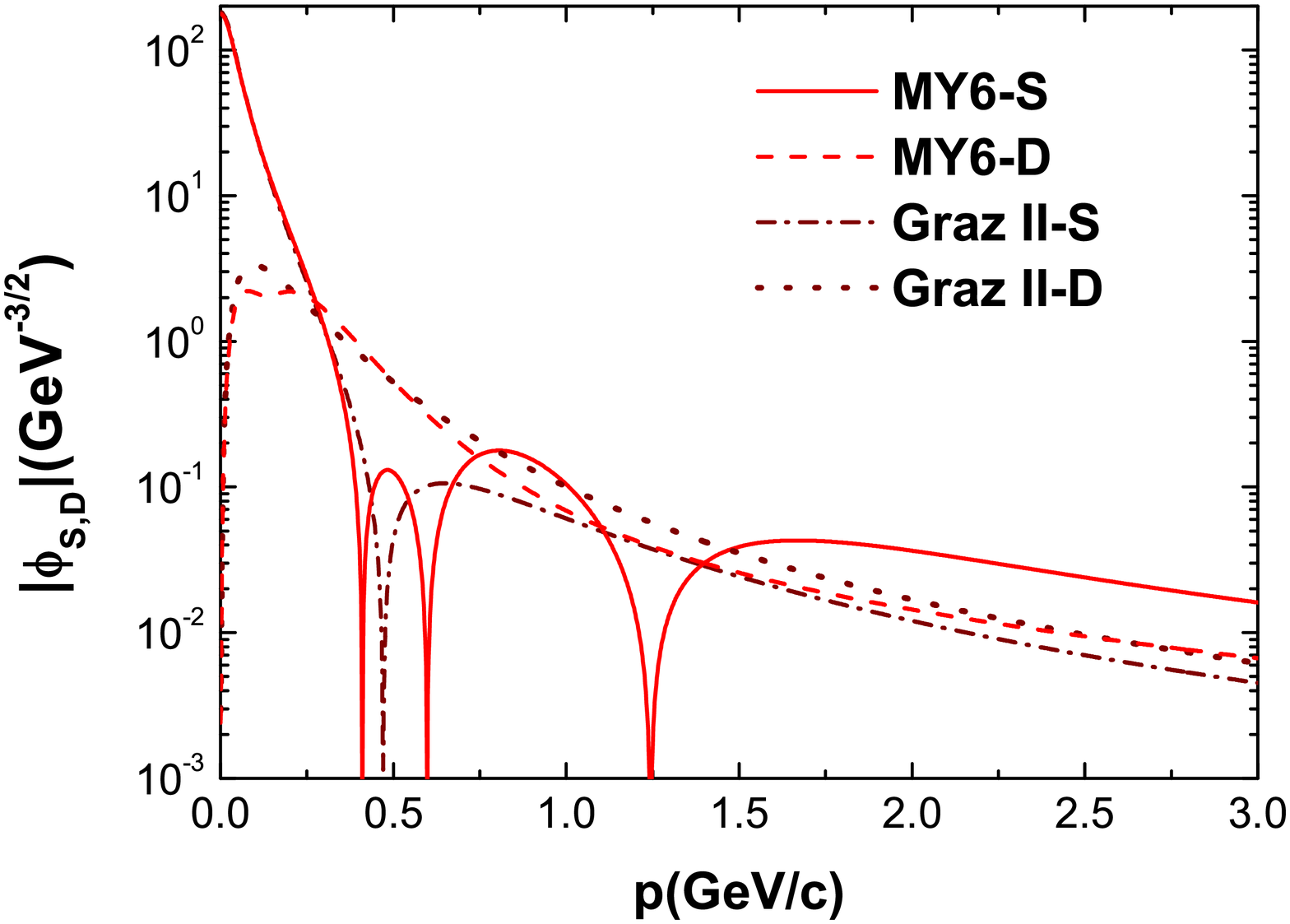}
\end{center}
\caption{The radial parts of the amplitude (\ref{amp_vf_d}) for
the $\Spp$ and $\Dpp$ partial-wave states at $ k_0=M_d/2-E_\bk$
are presented. They are written in the deuteron rest frame. The
MY6 model \cite{new} (MY6-S red solid line corresponds to $\Spp$
partial-wave state, MY6-D red dashed line - to $\Dpp$) is compared
with Graz II \cite{Rupp:1989sg} (Graz II-S brown dash-dotted line
- $\Spp$ wave function, Graz II-D brown dotted line - $\Dpp$ wave
function).} \label{fig:7}
\end{figure}

\begin{figure}
\begin{center}
  \includegraphics[width=0.55\textwidth]{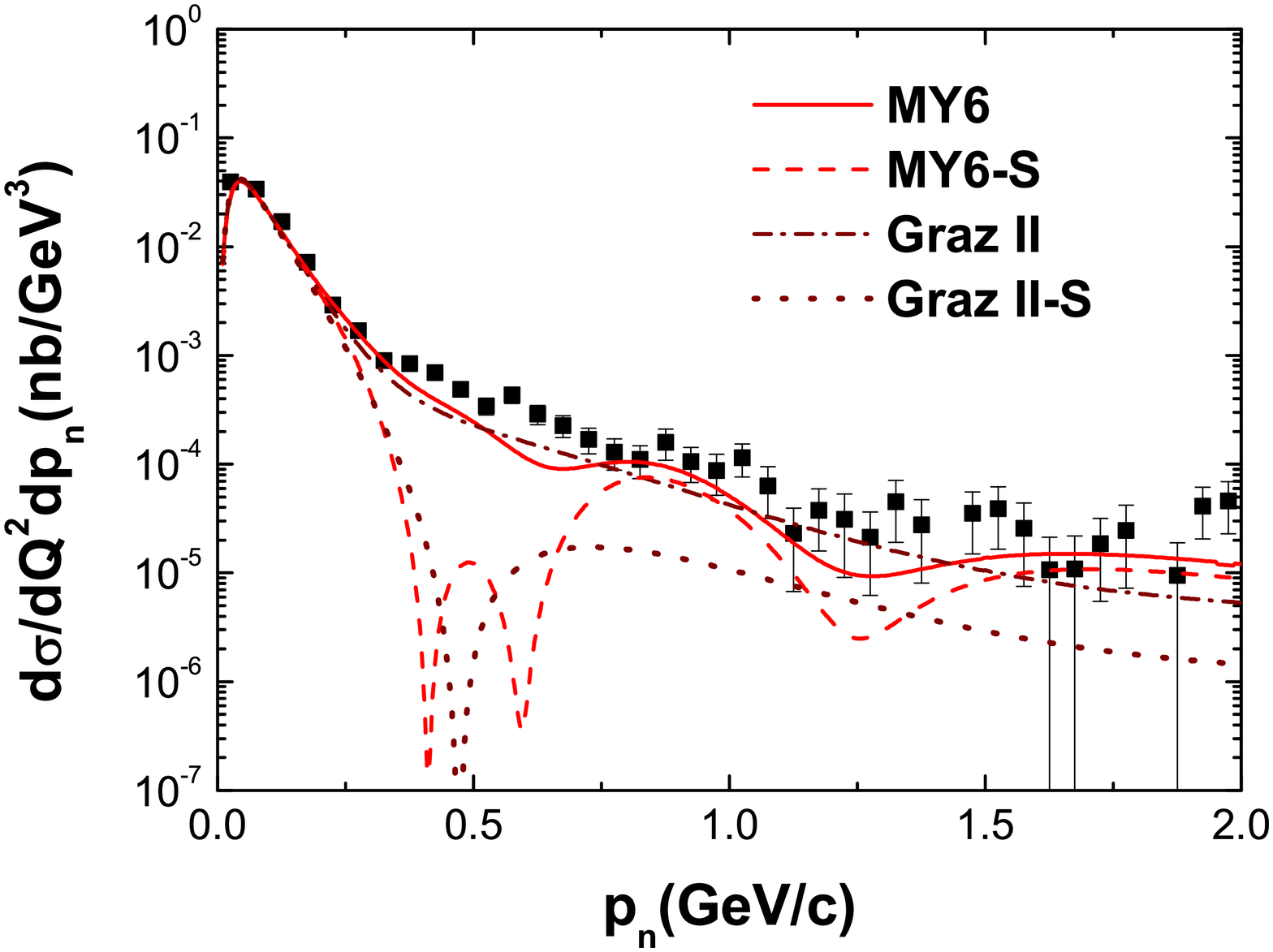}
\end{center}
\caption{The contribution of the $\Spp$ partial-wave state to the
cross section (\ref{cs_jlab}) is shown. Calculations with
(MY6 red solid and Graz II brown dash-dotted lines) and without the
$\Dpp$ state (MY6-S red dashed and Graz II-S brown dotted lines)
in the deuteron are compared at $Q^2=5\pm0.5$\,GeV$^2$.}
\label{fig:8}
\end{figure}

It should be noted that the behavior of the calculated cross
section is similar to the behavior of the corresponding wave
function for the deuteron $\Spp$ partial-wave state which is shown
in Fig.\ref{fig:7}. It is seen from the comparison of the cross sections
calculated with (MY6, Graz II) and without $\Dpp$ partial-wave
state (MY6-S, Graz II-S) in the deuteron (Fig.\ref{fig:8}) that the
influence of the $\Spp$ state is maximum at low and medium neutron
momenta. The minima of the $^3S_1^+$ wave functions are noticeable
in the cross section. They are smoothed by the $\Dpp$
state when the both partial-wave states are taken into account. The
role of the $\Dpp$ state increases at high $\bp_n$.

It is seen from Figs.\ref{fig:1}-\ref{fig:6} that the Graz II
and MY6 models give qualitatively the similar description of the
experimental data within the used approximation.
The difference between the model calculations and the experimental
data can be probably eliminated when FSI, $P$ waves and TBC are
taken into account. It should be emphasized that there is no
possibility to improve a theoretical description using the Graz II
model. As it was mentioned above, FSI is impossible to calculate
with the Graz II kernel for high-energy particles unless the $p_0$
component is constrained by some assumption like, for instance, in
quasipotential ap\-pro\-ac\-hes. On the contrary, it is possible
to take FSI into account using the MY6 kernel without constraining
$p_0$.

We should also comment the results obtained using the Paris
potential model \cite{Lacombe:1980dr} in \cite{Egiyan:2007qj}. A
good agreement with the experimental data was achieved when the
FSI and meson-exchange current effects were taken into account.
However, it seems questionable to use the nonrelativistic
potential elaborated to describe the $np$ elastic scattering data
for laboratory energies of the colliding particles less than
350\,MeV when the cross section at high $Q^2$ and $|\bp_n|$ is
calculated. The MY6 model has two important advantages. Firstly,
it is fitted to describe all available elastic $np$ scattering
data \cite{new}. Secondly, it allows one to perform calculations
without any necessity to constrain $p_0$
\cite{Bondarenko:2008ha}-\cite{new}.

In this paper the comparison of three different models of NN
interactions has demonstrated that relativistic effects play an important role in the description of the deuteron
electrodisintegration at high momentum transfer. The result is slightly dependent of the model
used for the proton and neutron electromagnetic form factors.
Further investigation is required to conclude which
models of NN interactions and nucleon electromagnetic form factors
are reasonable. In particular, it is necessary to calculate
FSI, $P$ waves and so on.

%
%
\begin{acknowledgements}
Authors would like to thank the organizers of "Relativistic
Description of Two- and Three-Body Systems in Nuclear Physics"\
Workshop for an opportunity to present and discuss this work.
Speaker R.E.P. is grateful to them for their warm hospitality.
\end{acknowledgements}
%


\end{document}